\documentclass[12pt,a4paper]{article}
\usepackage{amssymb}
\def\t#1{\tilde #1}
\def\ad{\mbox{ad}\,}
\def\b#1{{\mathbb #1}}

\def\c#1{{\cal #1}}

\def\Dirac{{\raise0.09em\hbox{/}}\kern-0.69em D}

\def\kbar {{\mathchar'26\mkern-9muk}}

\def\ep{i\epsilon}

\def\pprime{{\prime\prime}}
\def\t#1{\tilde #1}
\def\tfrac #1#2{\textstyle{\frac{#1}{#2}}}
\def\dfrac #1#2{\displaystyle{\frac{#1}{#2}}}
 
\def\k{\kern-.1em\mathbin{,}\kern-.1em}
\def\hk{\kern.12em\raise-1em\hbox{$\hat{\raise1em\hbox{,}}$}\kern.12em}
\newcommand{\initiate}{\setcounter{equation}{0}}
\def\be{\begin{equation}} \def\bea{\begin{eqnarray}}
\def\ee{\end{equation}}\def\eea{\end{eqnarray}}
\def\ncr{\nonumber\\ }

\begin{document}

\title{Noncommutative 2-Dimensional Models of Gravity}

\author{M. Buri\'c$\strut^1$,   J. Madore$\strut^{2}$ \\
      $\strut^{1}$Faculty of Physics, P.O. Box  368\\
      11001 Belgrade
\and    $\strut^{2}$Laboratoire de Physique Th\'eorique\\
      Universit\'e de Paris-Sud, B\^atiment 211, F-91405 Orsay}

\date{}

\maketitle

\abstract{A review is given of some 2-dimensional metrics for which
noncommutative versions have been found. They serve partially to
illustrate a noncommutative extension of the moving-frame
formalism. All of these models suggest that there is an
intimate relation between noncommutative geometry on the one hand and
classical gravity on the other.}

\vfill

\noindent ESI preprint 1467 \vspace{1cm}


 \eject

\parskip 4pt plus2pt minus2pt

\initiate
\section{Introduction}

There is a very simple argument due to Pauli that the quantum effects
of a gravitational field will in general lead to an uncertainty in the
measurement of space coordinates. It is based on the observation that
two `points' on a quantized curved manifold can never be considered as
having a purely space-like separation. If indeed they had so in the
limit for infinite values of the Planck mass, then at finite values
they would acquire for `short time intervals' a time-like separation
because of the fluctuations of the light cone.  Since the `points' are
in fact a set of four coordinates, that is scalar fields, they would
not then commute as operators. This effect could be considered
important at least at distances of the order of Planck length, and
perhaps greater. This is one motivation to study noncommutative
geometry. A second motivation, which is the one we consider ours, is
the fact that it is possible to study noncommutative differential
geometry, and there is no reason to assume that even classically
coordinates commute at all length scales. One can consider for example
coordinates as order parameters as in solid-state physics and suppose
that singularities in the gravitational field become analogs of core
regions; one must go beyond the classical approximation to describe
them.  A straightforward and conservative way to do this is to
represent them by operators. The space-time manifold is thus replaced
with an algebra $\c{A}$ (noncommutative `space'), generated by a set
of noncommutative `coordinates' $x^i$. We think of $x^i$ as linear
operators on some vector space, and therefore we assume that the
multiplication in the algebra is associative. The essential element
which allows us to interpret a noncommutative algebra as a space-time
is the possibility~\cite{Wor89,Con94} to introduce a differential
structure on the former.

We use a noncommutative version of Cartan's frame
formalism~\cite{Mad00c}; the differential structure has been also
studied from other points of view~\cite{Con94,Lan97,FigGraVar00}.
In order to develop some intuition in the complete absence of
experimental evidence, one is obliged to consider examples.
Several of these have been found.  We shall introduce as
illustrations the quantum plane, the 2-dimensional de~Sitter
space, the 2-dimensional Rindler space and the `fuzzy donut'.
Their simplicity will allow us to bypass the general
formalism and will permit a more intuitive presentation. A series
of models in all dimensions has been found~\cite{CerFioMad00a}, as
well as some models in dimensions two~\cite{Mad00c,JamSyk04} and
four~\cite{BurMacMad04,DimJomMolTsoWesWoh03}.  We shall argue that
the moving frame formalism is in this respect a natural way to
implement gravity.  It enables one to introduce a sort of
correspondence principle as a guide of how to construct the frame
from its commutative limit.  Some of our examples have been
considered elsewhere; we believe the Rindler and the donut examples
 to be new.  We  discuss their properties in more detail in the
last  sections and therefore on a less introductory level. 
The parameter $\kbar$ with the dimensions of
length squared is introduced in Section~\ref{Lob} to facilitate
the discussion of the commutative limit; a mass parameter $\mu$ as
well as Newton's constant, are also introduced.

\initiate
\section{Differential calculi}

In the following sections we shall use some known examples to
introduce the reader to the more elementary aspects of the
noncommutative extension of the de~Rham differential calculus.  In
ordinary geometry a topological manifold can have more than one
differential structure. In the noncommutative case this is generically
so: many non-equivalent differential structures on a given algebra
exist. This means that there are several `geometries' which one can
associate to the underlying noncommutative space. Once the
differential calculus is chosen, some more or less obvious assumptions
as hermiticity and bilinearity fix the linear connection almost
uniquely. This is to be contrasted with the commutative case, where
for each differential structure linear connection can be chosen almost
arbitrarily. The choice therefore of a differential is one of the more
important steps in the `quantization'. We stress that we do not think
that all noncommutative geometries are suitable as noncommutative
models of space-time, any more so than in the commutative limit.

The outline of the presentation is the following. In
Section~\ref{Manin} we shall introduce derivations as the
noncommutative version of vector fields and in Section~\ref{Lob} we
shall consider more particularly inner derivations and introduce
momentum generators.  In Sections 5,6 and 7 we discuss the
nonequivalence of the calculi induced by different frames, the frame
rotations and the noncommutative limit.  In Section~\ref{Donut} we
show that the momenta close under the commutator to generate a
quadratic algebra.  Implicit in our calculations is the assumption
that this algebra can be at least formally identified with the
original algebra generated by the `coordinates'. Although not
explicitly mentioned, this algebra is present in most of the examples;
the Section~\ref{Rind} presents what could be considered a
counterexample with an algebra which is not quadratic. We conclude by
introducing connection, metric and curvature in the noncommutative
case in Section \ref{dg} and discussing their relations in Section
\ref{ho}.

\section{Quantum plane: differentials}             \label{Manin}

We introduce here the quantum plane to illustrate the basic
features of noncommutative geometry; more details can be found for
example in the books mentioned above. A noncommutative `space' is
an associative $*$-algebra $\c{A}$ generated by a set of hermitean
`coordinates' $x^i$ which in some limit tend to the (real)
coordinates $\t{x}^i$ of a manifold; the latter we identify as the
classical limit of the geometry. That is, in the classical or
weak-field limit we impose the condition
\begin{equation}
x^i \to Z^{-1}\t{x}^i.
\end{equation}
The $Z$ could be perhaps singular; in all examples considered here
we can choose $Z=1$.  Elements of $\c{A}$ will be denoted by $x^i$,
$f$, $g$, $p_a$, and so forth. In general the coordinates satisfy a
set of commutation relations. We shall consider the algebra as a
formal algebra and not attempt to represent it as an algebra of
operators.

The simplest relation which can be used  to define the algebra is
\begin{equation}
\left[ x^i,x^j\right] = iJ^{ij},                   \label{can}
\end{equation}
where $J^{ij}$ are real numbers defining a canonical or symplectic
structure. Often in the literature the notation $\theta^{ij}$ is
used instead of $J^{ij}$. Another associative algebra is defined
using the commutation relations
\begin{equation}
\left[ x^i,x^j \right] = i C ^{ij}{}_{ k} x^k. \label{Lie}
\end{equation}
For simplicity we shall assume that the center of the algebra, the set
of elements which commute with all generators, consists of complex
multiples of the identity. The third important special case is a quantum
space, defined by a homogeneous quadratic relation:
\begin{equation}
\left[x^i, x^j\right] = iC^{ij}{}_{kl} x^kx^l. \label{qspace}
\end{equation}
A combination of these three commutation relations will be
satisfied by a set of generators $p_a$ of $\c{A}$ which we shall
refer to as momenta; the commutation relations obeyed by the
`coordinates' in general are not even necessarily polynomial.

In ordinary geometry a vector field can be defined as a derivation
of the algebra of smooth functions. This definition can be used
also when the algebra is noncommutative. A derivation, we recall,
is a linear map $f\mapsto Xf$ which satisfies the Leibniz rule,
$X(fg) = (Xf) g+f Xg$. Derivations will be denoted by $X$, $Y$,
$e_a$ and so forth and the set of all derivations by
$\mbox{Der}(\c{A})$. A simple example is the algebra of $2\times
2$ complex matrices $M_2$ with (redundant) generators the Pauli
matrices. The algebra is of dimension four, the center is of
dimension one and $\mbox{Der}(M_2)$ is of dimension three with
basis consisting of three derivations $e_a=\ad \sigma_a$:
\begin{equation}
e_af = [\sigma_a,f].
\end{equation}
We notice that the Leibniz rule is here the
Jacobi identity. We see also that the left multiplication
$\sigma_ae_b$ of the derivation $e_b$ by the generator $\sigma_a$
no longer satisfies the Jacobi identity: it is not a derivation.
The vector space $\mbox{Der}(M_2)$ is not a left $M_2$ module.
This property is generic. If $X$ is a derivation of an algebra
$\c{A}$ and $h$ an element of $\c{A}$, then $hX$ is not
necessarily a derivation:
\begin{equation}
hX(fg) = h (Xf) g +h f Xg \neq h(Xf) g+f hXg
\end{equation}
if $hf \neq fh$. Notice that the derivations of $M_2$ are
inner: they were defined as a commutator with an element of the
algebra. It is a simple theorem that all derivations of a complete
matrix algebra are inner. On the other hand, the derivations on an
algebra of functions are not inner; they are known as outer.

Although derivations do not form a left module, one can introduce
associated elements known as differential forms which form a bimodule;
they can be multiplied from the left and from the right. We shall
therefore express as much as possible physical quantities using the
latter. We define here a 1-form $\omega$ is a linear map $\omega :
\mbox{Der}(\c{A})\to \c{A}$. The set of 1-forms $\Omega ^1(\c{A})$ has
a bimodule structure, that is, if $\omega$ is a 1-form, $f\omega$ and
$\omega f$ are also 1-forms.  The elements of $\Omega ^1(\c{A})$ will
be typically denoted by $\omega$, $\theta$, $\xi$, $\eta$.

The important step is the definition of a differential $d$; it is
a linear map from functions to 1-forms, $d: \c{A} \to \Omega
^1(\c{A})$ which obeys the Leibniz rule. In general $fdg\neq dgf$
but we shall introduce later special forms $\theta^a$ which
commute with the algebra. The exterior product $\xi\and\eta$ of
two 1-forms $ \xi$ and $\eta$ is a 2-form. There is no reason to
assume the exterior product antisymmetric. We mention also that
one can deduce the structure of the algebra of all forms from that
`of the module of 1-forms.  The map $d$ can be extended to  all
forms if one require that $d^2 = 0$. We should stress that in
general one can associate  many differential calculi to a given
algebra.

To illustrate these notions, we construct a differential for the
canonical structure (\ref{can}): $\left[ x^i, x^j\right] =
iJ^{ij}$.  From the Leibniz rule it follows that the differential
of the unit element must vanish. Therefore the differential must
satisfy the constraint
\begin{equation}
0 = d\left[ x^i, x^j\right] = dx^i x^j+x^i dx^j -dx^j x^i -x^j
dx^i.
\end{equation}
A possible but not unique solution to this equation is
\begin{equation}
dx^i x^j-x^j dx^i = \left[ dx^i,x^j\right] = 0.
\end{equation}
Furthermore, the relations of the algebra imply that
\begin{equation}
0 = d\left[ dx^i,x^j\right] = (d^2x^i) x^j -dx^i dx^j-dx^j dx^i
+x^j (d^2x^i),
\end{equation}
that is, the differentials anticommute as if they were defined on
a manifold,
\begin{equation}
dx^i dx^j = - dx^j dx^i.
\end{equation}
For the Lie algebra~(\ref{Lie}) however, we see that we could not
have imposed the condition $ \left[ dx^i,x^j\right] = 0$ as it is
inconsistent with the relation $\left[ x^i,x^j\right] =
iC^{ij}{}_{ k}x^k$. It would imply $C^{ij}{}_{k} dx^k = 0$.

The first example we discuss in detail is the quantum plane. It has
two generators $x$ and $y$ related by
\begin{equation}
xy = qyx,                                        \label{qplane}
\end{equation}
where $q$ is a constant which we shall assume not to be a root of
unity. For example, two derivations $e_1$, $e_2$ can be defined by the
formulae
\begin{equation}
\begin{array}{ll} e_1 x = x, & \quad  e_2 x = 0, \\[6pt] e_1 y
= 0, & \quad e_2 y = y.
\end{array}
\end{equation}
These would necessarily be outer derivations.  There are other
possibilities.  Let $e_a$ be defined by
\begin{equation}
\begin{array}{ll}
e_1 x = \displaystyle{\frac{q^2}{q^2+1}} x^{-1}y^2, &\quad
e_1 y = \displaystyle{\frac{q^4}{(q^2+1)}} x^{-2}y^3, \\[6pt]
e_2 x = 0, &\quad
e_2 y = -\displaystyle{\frac{q^2}{(q^2+1)}} x^{-2}y.
\end{array}                                                 \label{name}
\end{equation}
These derivations are, as we shall see, inner.

A differential must satisfy the constraint
\begin{equation}
d(xy-qyx) = dx \,y +x\, dy - q\, dy\, x- q y\, dx = (dx\, y-q\,
dy\, x)+(x\, dy - q y \,dx) = 0.
\end{equation}
This we can satisfy, for example,
by setting
\begin{equation}
dx\, y - q y \,dx =0, \quad  q  dy\, x-x\, dy =
0,\label{R1}
\end{equation} which defines the commutation rules of
$dx$ and $ y$, and $dy$ and $ x$. In order to complete the
definition one must add the  rules for  $dx$ and $ x$ and $dy$ and
$ y$. For example,
\begin{eqnarray} dx y = q dy x, \quad &&
x \,dy = q\,  dy \, x,                            \label{QP1}\ncr
x\, dx = q\, dx\, x, \quad
&& q y\, dy = dy\, y.
\end{eqnarray}
Applying $d$ once more, one thus obtains the exterior product
\begin{equation}
(dx)^2 = 0,\quad ( dy)^2 = 0,\quad dx\, dy = -q \, dy\, dx.\label{QP11}
\end{equation}

For reasons \cite{WesZum90} which do not concern us here (deformed
symmetries), one prefers
another differential calculus constructed by setting instead
of~(\ref{R1})
\begin{equation}
dx\, y -  q y\, dx  = (1- q^2) dx\, y  , \quad q\, dy x -x\, dy =
(1-q^2) dx \, y.                                       \label{R2}
\end{equation}
The full set of relations for the 1-differential forms would be in
this case
\begin{equation}
\begin{array}{ll}
q\, dx\, y = y\, dx,&
\quad  x\, dy = q\, dy x+(q^2-1) dx\, y,\\[6pt]
x\, dx = q^2 dx\, x,& \quad y\, dy = q^2 dy\, y,
\end{array}                                               \label{QP2}
\end{equation}
In this case  the exterior product is given by
\begin{equation}
  (dx)^2 = 0,\quad ( dy)^2 = 0,\quad q\, dx\, dy = - dy\, dx.
\end{equation}
We shall see that it is based on the inner derivations (\ref{name})
defined above. The relation between the unusual structure of these
derivations and the deformed symmetries is not completely understood.

\initiate
\section{De Sitter: frames}                      \label{Lob}

As we have seen, there is a variety of possibilities to define a
differential. One problem is how to determine or at least restrict it
by imposing some physical requirements. We shall use here a
modification of the moving frame formalism and show that so defined
differential calculi over an algebra admit essentially a unique metric
and linear connection. We shall fix therefore the differential
calculus by requiring that the metric have the desired classical
limit.  The idea is to define an analogue of a parallelizable
manifold, which therefore has a globally defined frame. The frame is
defined either as a set of vector fields $e_a$ or as a set of 1-forms
$\theta^a$ dual to them. The metric components with respect to the
frame are then constant.

We choose a set of $n$ derivations $e_a$ which we assume to be
inner generated by  `momenta' $p_a$:
\begin{equation}
e_af = [ p_a,f].                               \label{ep}
\end{equation}
We
suppose that the momenta generate also the whole algebra $\c{A}$. Since
the center is trivial, this means that an element which commutes
with all momenta must be a complex number. An alternative way is to
use the 1-forms $\theta^a$ dual to $e_a$ such that relation
\begin{equation}
\theta^a(e_b) = \delta^a_b                          \label{jmmb}
\end{equation}
holds. To define the left hand side of this equation we define first
the differential, exactly as in the classical case, by the condition
\begin{equation}
df(e_a) = e_a f.                    \label{diff}
\end{equation}
The left and right multiplication by elements of the algebra
$\c{A}$ are defined by
\begin{equation}
f dg = f e_ag \theta^a  ,\qquad                      \label{bi1}
dg f = e_ag f \theta^a.
\end{equation}
Since every 1-form can be written as sum of such terms the definition
is complete. In particular, since
\begin{equation}
f \theta^a(e_b) = f\delta^a_b = (\theta^a  f)(e_b),   \label{cond}
\end{equation}
we conclude that the frame necessarily commutes with all the elements
of the algebra $\c{A}$; this is a characteristic feature.  If one does
not insist on using differential calculi defined by inner derivations
this condition can be generalized to include frames which commute only
modulo an algebra morphism. For a recent discussion of this
possibility we refer to~\cite{DimMue04}.

In the case of the algebra $M_2$ considered above, the module of
1-forms is generated by three elements $d\sigma_a$ defined as the maps
\begin{equation} 
d\sigma_a(e_b) = e_b \sigma_a = [\sigma_b,\sigma_a]. \end{equation}
The maps $\sigma_c d\sigma_a$ and $d\sigma_a\sigma_c$ are defined
respectively as
\begin{equation} 
\sigma_cd\sigma_a(e_b)  = \sigma_c
[\sigma_b,\sigma_a],\qquad d\sigma_a \sigma_c(e_b) =
[\sigma_b,\sigma_a]\sigma_c.
\end{equation} 
Obviously, $\sigma _c d\sigma_a\neq d\sigma_a\sigma_c$.

The 1-form $\theta$ defined as
\begin{equation}
\theta = -p_a\theta^a
\end{equation}
can be considered as an analog of the Dirac operator in ordinary
geometry.  It implements the action of the exterior derivative on
elements of the algebra. That is
\begin{equation}
df =  -\left[ \theta,f\right] = 
\left[p_a\theta^a,f\right] =
\left[p_a,f\right] \theta^a.  
\end{equation}
The differential is real if $(df)^* = df^*$.  This is assured if the
derivations $e_a$ are real: $e_af^* = (e_a f)^* $, which is the case
if the momenta $p_a$ are antihermitean. From the definitions one has
${\theta^a}^* = \theta^a$, $ \theta^* = -\theta $. Furthermore, 
$(f\xi)^* = \xi^* f^*$, $ (\xi f)^* = f^*\xi^*$, and 
$(\xi\eta )^* = -\eta^*\xi^* $. Note that whereas the product of two
hermitean elements is hermitean only if they commute, the product of
two hermitean 1-forms is hermitean only if they anticommute.

Consider once more the quantum plane introduced in the previous
section. The momenta $p_a$ can be defined as
\begin{equation}
p_1 = \frac{q}{q-1}\, y,\quad p_2 =  \frac{q}{q-1} \, x.
\end{equation}
From these expressions one easily finds the relations
\begin{equation}
\begin{array}{ll}
e_1x = -xy, &\quad e_2x=0, \\[4pt]
e_1 y = 0, & \quad  e_2y = xy.
\end{array}
\end{equation}
Using the definition $df = e_a(f) \theta^a$, one obtains
for $\theta^1$ and $\theta^2$,
\begin{equation}
\theta^1 = -y^{-1}x^{-1}dx,\quad  \theta^2 = y^{-1}x^{-1}dy.
\end{equation}
From (\ref{cond}) the module structure (\ref{QP1}) can be
reconstructed. The momenta $p_a$ satisfy the quadratic relation
\begin{equation}
p_2p_1 = q p_1p_2.                                \label{1}
\end{equation}
The second differential calculus~\cite{WesZum90} on the quantum
plane also has a frame. The corresponding momentum generators are
\begin{equation}
p_1 = \frac{1}{q^4-1}\, x^{-2}y^2,\quad p_2 = \frac{1}{q^4-1}\,
x^{-2}.
\end{equation}
They satisfy
\begin{equation}
p_1p_2 = q^4 p_2p_1.                                     \label{2}
\end{equation}
Note that the momenta are singular in the limit $q\to 1$. In quantum
mechanics the relation between the differential and the momentum $p$ is
given by
\begin{equation}
\frac{\partial f}{\partial x} =\frac{i}{\hbar}\, [p,f],
\end{equation}
whereas it is given here by the expression~(\ref{ep}). The
singularity of the classical limit $\hbar\to 0$ has been included in
the definition of the momentum.

The implementation of the differential structure as we have given
is just as arbitrary as before since it amounts to a choice of the
momenta. In some cases, the construction of the frame is not
difficult. In the example (\ref{can}) one can choose the
differential such that $\left[ x^i, dx^j\right] = 0$; a frame is
$\theta^a = \delta^a_i dx^i$ since $dx^i$ commute with all
elements of the algebra. The most general form is $\theta^a =
\Lambda^a_idx^i$ with $\Lambda^a_i$ real numbers. The duality
relations give the momenta
\begin{equation}
\delta^a_b = \theta^a(e_b) = \delta^a_i dx^i(e_b) = \delta^a_i
e_b(x^i) = \delta^a_i  [p_b,x^i],
\end{equation}
that is,
\begin{equation}
p_b = -i \delta^k_b J^{-1}_{ki} x^i.
\end{equation}
In  order to discuss noncommutative limit, (\ref{can})
should in fact be rewritten as
\begin{equation}
[x^i,x^j] = i\kbar J^{ij},
\end{equation}
introducing the parameter $\kbar$ to describe the fundamental area scale on
which noncommutativity becomes important. The $\kbar$ is presumably of order of
the Planck area $G\hbar$; the commutative limit is  defined by
$\kbar\to 0$. The momenta read then
\begin{equation}
p_a = \frac{1}{i\kbar} J_{ai}^{-1}x^i , \label{mom}
\end{equation}
and they are singular in the limit $\kbar\to 0$. 

Since the frame
is given by $\theta^a = \delta^a_i dx^i$, this space can be
naturally thought of as the noncommutative generalization of flat
space. The momenta are linear in the coordinates and hence
\begin{equation}
\left[p_a,p_b\right] =
\frac {1}{(i\kbar)^2}J^{-1}_{ai}J^{-1}_{bj}\left[ x^i,x^j\right] =
K_{ab}, \qquad K_{ab} = -\frac 1{i\kbar} J^{-1}_{ab}.
\end{equation}

In general only by explicit construction can one show that the frame
exists.  In the case of the Lie algebra (\ref{Lie}), for example, one
sees that the 1-forms $dx^i$ do not define a frame because they do not
commute with the algebra.  In the example of $M_2$ with Pauli matrices
as momenta, the frame which is the solution to the
equation~(\ref{jmmb}) is seen to be
\begin{equation}
\theta^a = \tfrac 14\sigma_b\sigma^a d\sigma^b.
\end{equation}
This construction can be repeated~\cite{DubKerMad89b} for the algebra
$M_n$ of $n\times n$ complex matrices.

As the main example of this section we consider the algebra
generated by two hermitean elements $x$ and $y$ related by
\begin{equation}
[x,y] = -2i\kbar\mu y.                                     \label{AdS}
\end{equation}
This is related to the Jordanian deformation~\cite{Agh93} of $\b{R}^2$
with deformation parameter $h = i\kbar\mu^2$. The $\mu$ is the
gravitational mass scale; the associated length $G\mu$ vanishes with
$\kbar$.  To find the frame, we rewrite this as follows
\begin{equation}
(x+i\kbar\mu ) y = y (x-i\kbar\mu ).
\end{equation}
The differential must satisfy
\begin{equation}
dx\, y +(x+i\kbar\mu )\, dy = dy\, (x-i\kbar\mu ) + y\, dx.
\end{equation}
We shall impose separately the conditions
\begin{equation}
dx\, y = y\, dx,\qquad (x+i\kbar\mu )\, dy = dy\, (x-i\kbar\mu ).
\label{I}
\end{equation}
The first of these relations  suggests that $dx$ can be taken as a
frame element, in fact  $f(y) dx$ as well as $dx$. We set
$\theta^1 = f(y) dx$. Rewriting~(\ref{I}) as
\begin{equation}
(x+i\kbar\mu ) y y^{-1}\, dy = y y^{-1}\, dy\, y^{-1} y
(x-i\kbar\mu ),
\end{equation}
we see that we can take $\theta^2 = -(\mu y)^{-1} dy$. We
assume that $\theta^a$ commute with  $x$ and $y$. The duality
relations (\ref{jmmb}) determine $f(y)$. They read
\begin{equation}
\begin{array}{ll}
f(y) [ p_1,x] = 1, &\quad f(y) [p_2,x] = 0, \\[4pt]
(\mu y)^{-1}[p_1,y] = 0,&\quad (\mu y)^{-1}[p_2,y] = -1,
\end{array}
\end{equation}
and reduce to (\ref{I}) for
\begin{equation}
p_2 = \frac{1}{2i\kbar}\, x,\qquad p_1 = \frac{1}{2i\kbar} \,
y,\qquad f(y) = (\mu y)^{-1}.
\end{equation}
The frame therefore is given by
\begin{equation}
\theta^1 = (\mu y)^{-1} dx,\quad \theta^2 = -(\mu y)^{-1} dy.
\end{equation}
The corresponding calculus is the covariant one~\cite{Agh93}.
The momenta are proportional to the coordinates
so their commutation relation is:
\begin{equation}
\left[p_1,p_2\right] = \mu p_1.
\end{equation}
A short calculation shows that the frame elements anticommute.
The line element
\begin{equation}
ds^2 = \pm (\t{\theta^1})^2 \pm (\t{\theta^2})^2
\end{equation}
of the commutative limit is that of the Lobachevski plane or of
(anti) de~Sitter space depending on the choice of signature.

\initiate
\section{Fuzzy sphere:  generalized frames}   \label{Sph} 

We now present an example of a calculus which possesses a frame
but not one which is dual to a set of derivations. It is a 2-dim
calculus which resembles a noncommutative version of the de~Rham
calculus over the 2-sphere. As such it can of course have no
commutative limit.  Consider an algebra with two generators
$\chi$, $\phi$ and introduce a frame
\begin{equation}
\theta ^1 = d\chi,\qquad \theta^2 = \sin\chi \,d\phi.
\end{equation}
The parameter $\mu$ is the inverse radius of the sphere and we set
$\epsilon = \mu^2\kbar$. As usual we denote $[\chi ,\phi] = \ep
J^{12}$ and assume that the frame elements commute with the
generators $\chi$, $\phi$. This gives immediately the relations
\begin{equation}
[\chi ,d\phi ] = 0,\quad
[\chi ,d\chi ] = 0,\quad
[\phi ,d\chi ] = 0 ,                 \label{1a}
\end{equation}
which imply that
\begin{equation}
(\theta^1)^2 = 0,\quad
\theta^1\theta^2 = -\theta^2\theta^1,
\end{equation}
as well as
\begin{equation}
dJ^{12} = 0.              
\end{equation}
We set $J^{12}=1$. The remaining relations yield the identity
\begin{equation}
[\phi, d\phi] = \ep \cot \chi\,d\phi.
\end{equation}
We take the differential to obtain
\begin{equation}
(\theta^2)^2 = - \ep \,\dfrac{1}{2\sin \chi}\,\theta^1\theta^2.
\end{equation}
We have defined  a differential calculus with the differential
\begin{equation}
d\theta^1 = 0, \quad d\theta^2 = \cot\chi \,\theta^1 \theta^2,
\end{equation}
but as we shall now
show not one which is dual to a set of inner derivations.

The relations
\begin{equation}
\begin{array}{ll}
[p_1, \phi] = 0, & [p_1, \chi] = 1,\\[2pt]
[p_2, \phi] = \dfrac{1}{\sin \chi} , & [p_2, \chi] = 0
\end{array}
\end{equation}
define the duality between momenta and coordinate generators.
These can be solved to yield
\begin{equation}
\ep p_1 = - \phi, \qquad
\ep p_2 =  G(\chi),             \label{ime}
\end{equation}
with the function $G(\chi)$ defined by
\begin{equation}
\dfrac{dG}{d\chi} = \dfrac{1}{\sin \chi},  \quad
G(\chi) = \log\tan\frac \chi 2 .
\end{equation}
Using these momenta we define a `Dirac operator'
\begin{equation}
\theta = - p_a \theta^a.
\end{equation}
From~(\ref{ime}) we find that $\theta$ is given by
\begin{equation}
\theta = \dfrac{1}{\ep} \phi d\chi -
\dfrac{1}{\ep}G(\chi )\sin \chi d\phi.
\end{equation}
We have then
\begin{equation}
d\theta + \theta^2 = - \dfrac{1}{\ep}\dfrac{1}{2\sin\chi}
(4 + 2G \cos \chi + G^2) \theta^1 \theta^2.               \label{bm}
\end{equation}
As we shall see later~(Equation~(\ref{d2})) the commutator of $f$
with~(\ref{bm}) defines the second differential $d^2 f$ of an
element $f$. Therefore it must commute with all elements of the
algebra; this is obviously not the case with~(\ref{bm}). Thus we
see that the differential defined by the set of
momenta~(\ref{ime}) is inconsistent.

A solution~\cite{ConLot92,Mad00c} to this problem is to consider
only 2-forms modulo the image of $d^2$. This would result in a
consistent differential calculus but with 2-forms depending only
on $\phi$, with perhaps special functions of $\chi$.  One should
not of course be too attached to the condition $d^2=0$. A
non-vanishing value for this operator could be interpreted as some
sort of `micro-curvature'. In a subsequent article~\cite{BurMad05}
the authors will examine the relation between this
`micro-curvature' and ordinary curvature using the WKB
approximation.

\initiate
\section{Rindler space: frame rotations}   \label{Rind}

In the last example of the previous section we saw that typically
there is little freedom in finding the solution to duality and
consistency equations. This is due to the relations among the momentum
generators which we shall derive in Section~\ref{Donut}. We shall
see there that this relation is at most quadratic.  As
an example of a frame for which such a noncommutative extension does
not exist we consider the 2-dim Rindler frame which is defined in
one-half of 2-dim Minkowski space.  We shall use this example to
illustrate the fact that not all moving frames are suitable for
`quantization'; some are more suitable than others.

Let $\mu$ be a parameter with dimensions of mass and proportional to
the Rindler acceleration.  The commutative Rindler frame is given by
$\t{\theta}^0 = \mu \t{x}d\t{t}$ and $\t{\theta}^1 = d\t{x}$; the
commutative Minkowski frame is $\t{\theta}^{\prime 0} = d\t{t}^\prime$
and $\t{\theta}^{\prime 1} = d\t{x}^\prime$. The local Lorentz
rotation from the former to the latter is defined by
\begin{equation}
\t{\theta}^{\prime a} = \t{\Lambda}^{-1}{}^a_b \t{\theta}^b \label{fr}
\end{equation}
with
\begin{equation}
\t{\Lambda} =
\left(\begin{array}{cc}
\cosh \mu \t{t} & \sinh\mu \t{t}  \\[2pt]
\sinh \mu \t{t} & \cosh \mu \t{t}
\end{array}\right).
\end{equation}
The classical coordinate transformation from the Rindler coordinates
to the Minkowski coordinates is given, for $x > 0$ by
\begin{equation}
\t{x}^\prime = \t{x} \cosh \mu \t{t}, \qquad
\t{t}^\prime = \t{x} \sinh \mu \t{t}.
\end{equation}
It is of course not to be confused with the rotation.

We shall first show that the Rindler frame is not a suitable frame; there
are no dual momenta. If momenta $p_a$ did exist then they
would necessarily satisfy the relations
\begin{equation}
\begin{array}{ll}
[p_{0}, t] = (\mu x)^{-1}, & [p_{0}, x] = 0, \\[2pt]
[p_{1}, t] = 0, & [p_{1}, x] = 1.
\end{array}                                                   \label{RF0}
\end{equation}
But one easily sees that the solution is not a quadratic algebra.
In fact if one set $[p_{0}, p_{1}] = L_{01}$ and
$[t,x] = i\kbar J^{01}$, one finds from the Jacobi identities that
\begin{eqnarray}
&&[L_{01}, t] =  [p_{0}, [p_{1}, t]] - [p_{1}, [p_{0}, t]]
= \mu^{-1} x^{-2}, \\[4pt]
&&[L_{01}, x] =  [p_{0}, [p_{1}, x]] - [p_{1}, [p_{0}, x]] = 0.
\end{eqnarray}
The $L_{01}$ commutes with $x$ and therefore belongs to the algebra
generated by $x$. From the commutation with $t$ one finds
\begin{equation}
i\kbar \mu  (\frac{d}{dx}L_{01}) J^{01} = - x^{-2}.
\end{equation}
Similarly one has
\begin{eqnarray}
&&i\kbar [p_{0}, J^{01}] =  [[p_{0}, t], x] - [[p_{0}, x], t] = 0,\\[4pt]
&&i\kbar [p_{1}, J^{01}] =  [[p_{1}, t], x] - [[p_{1}, x], t] = 0
\end{eqnarray}
from which one concludes that $J^{01}$ is constant. We shall set
$J^{01}=1$. We deduce therefore, neglecting integration constants, that
\begin{equation}
L_{01} = \frac{1}{i\kbar\mu} x^{-1}.
\end{equation}
But the duality relations~(\ref{RF0}) require
\begin{equation}
p_0 = -\frac{1}{i\kbar\mu}\log (\mu x), \qquad
p_1 = \frac{1}{i\kbar}\, t                          \label{p0x}
\end{equation}
and thus one easily sees that
\begin{equation}
L_{01} = \frac{1}{i\kbar } e^{i\kbar\mu  p_0}
\end{equation}
which is not a quadratic expression in $p_0$ and $p_1$.

The expressions~(\ref{p0x}) for the momenta seem quite different from
the corresponding commutative expressions for the derivations
$\t{e}_i$ dual to the frame:
\begin{equation}
\t{e}_0 = (\mu \t{x})^{-1} \partial_0, \qquad \t{e}_1 =
\partial_1.
\end{equation}
However in both cases one obtains the same action on the generators
of the algebra. In particular
\begin{equation}
\t{e}_0 \t{t} = (\mu \t{x})^{-1}, \qquad [p_0, t] = (\mu x)^{-1}.
\end{equation}
Here one appreciates the importance of the space-time commutation
relation.

Although the momenta $p_a$ dual to the frame which we have used do not
satisfy a quadratic relation it is easy to introduce another set
$\bar p_a$ which do. We define the new momenta by the equations
\begin{equation}
\bar p_0 = -\frac{1}{i\kbar\mu } e^{-i\kbar\mu p_0}, \qquad \bar
p_1 = p_1.                                           \label{?}
\end{equation}
They obey the commutation relation
\begin{equation}
[\bar p_0,\bar p_1] =  \dfrac 1{i\kbar}.
\end{equation}
From~(\ref{p0x}) one see also that $\bar p_a$ are related to the
coordinate generators by the transformations
\begin{equation}
x = - i\kbar \bar p_0, \qquad t = i \kbar  \bar p_1.
\end{equation}
The frame defined by the new momenta is given by $\theta^a =
\delta ^a_i \bar d x^i$; it is a Minkowski-like frame in Rindler
coordinates.  We put here a bar on the differential
to emphasize that the calculus is different. In spite of the
apparent nonlocality in the transformation~(\ref{?}) the action of
both $\bar p_0$ and $p_0$
\begin{equation}
[p_0, f] \to (\mu \t{x})^{-1}\t{\partial}_0 f, \qquad [\bar p_0,
f] \to \t{\partial}_0 f
\end{equation}
is local.

We now compare the differential calculi defined by two different
frames, Rindler and Minkowski, and related by a noncommutative
frame rotation of the type~(\ref{fr}).  Both frames can be used to
define a differential calculus; each differential calculus has at
most one basis as frame. Let $\t{\theta}^a$ be a global moving
frame for some 2-dimensional commutative geometry and let
$\{\t{\theta}^{\prime a}\}$ be the set of all moving frames
$\t{\theta}^{\prime a}$ such that
\begin{equation}
\t{\theta}^{\prime a} =
\t{\Lambda}^{-1}{}^a_b \t{\theta}^{b}
\end{equation}
for some local Lorentz rotation $\t{\Lambda}$. The set of
noncommutative versions will be described then each by a frame
$\theta^{\prime a}$ which we shall suppose related by
\begin{equation}
\theta^{\prime a} =
\Lambda^{-1}{}^a_b \theta^{b}
\end{equation}
for the corresponding `noncommutative' local Lorentz rotation. In
the special cases we have been considering one can restrict the
matrices $\Lambda$ to the subset the elements of which depend on
but one generator so they are well-defined. In more general
situations the definition would require elaboration. It is clear
that if $[f,\theta^{\prime a}] = 0$ then
\begin{equation}
[f,\theta^a] =
[f, \Lambda^a_b] \Lambda^{-1}{}^b_c \theta^c.   \label{tlo}
\end{equation}
This can also be written as a rule
\begin{equation}
\theta^a f =
\Lambda^a_b f \Lambda^{-1}{}^b_c \theta^c
\end{equation}
for relating the left- and right-module structures. In general
then each local rotation defines a different calculus. The
equivalence class of (commutative) moving frames gives rise to a
set of inequivalent frames which have the same classical limit. If
one wishes to consider one calculus, defined, say, by the
condition $[f,\theta^{a}] = 0$ then each of the bases
$\theta^{\prime a}$ satisfies the relation
\begin{equation}
\theta^{\prime a} f =
\Lambda^{-1}{}^a_b f \Lambda^b_c \theta^{\prime c}.
\end{equation}
That is, $\theta^{\prime a}$ is not the frame for the same
calculus unless $\Lambda$ is a global rotation (a constant
matrix).

Note that the Rindler metric can be considered equivalent to the
2-dim Kasner metric. The Kasner metric in dimension-4, for a
special value of the parameters, is flat and a moving frame can be
chosen which for these values become the ordinary flat frame:
\begin{equation}
\t{\theta}^0 = d\t{t}, \quad \t{\theta}^1 = d\t{x} - \t{t}^{-1}
\t{x} d\t{t},  \quad \theta ^2 = d\t{y}, \quad \theta^3 = d\t{z}.
\label{ckf}
\end{equation}
 We refer here to $\{ \t{\theta}^0,  \t{\theta}^1 \}$ as the
2-dim Kasner moving frame.  By a change of variables
\begin{equation}
\t{x} \to \t{t}, \qquad \t{t} \to {\t{t}}^{-1} \t{x}.
\end{equation}
the Rindler frame can be brought to this form. 

\initiate
\section{Kasner: noncommutative corrections}

We  shall here study the noncommutative corrections of  the 
Kasner metric as defined in (\ref{ckf}). For convenience instead
of $\t{x}$ and $\t{t}$ we choose as classical variables $\t{t}$
and
\begin{equation}
\t{\phi} = {\t{t}}^{-1} \t{x}.
\end{equation}
As we have already learned, not all moving frames attached to a
metric are suitable for quantization; in this case the appropriate
differential calculus is that determined by the flat Minkowski
frame. The classical frame rotation from the Minkowski moving
frame $\t{\theta}^a$ to the Kasner moving frame $\t{\eta}^a$ is
given by
\begin{equation}
\begin{array}{l}
\t{\eta}^{0} = \cosh\t{\phi} \, \t{\theta}^{0} - \sinh\t{\phi} \,
\t{\theta}^{1},
\\[2pt] \t{\eta}^{1} = - \sinh \t{\phi} \,
\t{\theta}^{0} + \cosh \t{\phi} \, \t{\theta}^{1}.
\end{array}
\end{equation}
The relation between the two coordinate systems is given by
\begin{equation}
\t{x}^\prime = \t{t} \sinh \t{\phi}, \qquad
\t{t}^\prime = \t{t} \cosh \t{\phi}.            \label{ttp}
\end{equation}
It follows that $\t{t}^2 = \t{t}^{\prime 2} -  \t{x}^{\prime 2}$; the
origin of the Kasner time coordinate, exactly at the flat-space values
of the parameters and because of the singular nature of the
transformation, becomes a null surface.

In the noncommutative case we choose the symmetric ordering;
therefore the change of generators~(\ref{ttp}) becomes
\begin{equation}
\begin{array}{l}
x^\prime = t \sinh \phi - \tfrac 12 [t, \sinh \phi], \\[4pt]
t^\prime = t \cosh \phi - \tfrac 12 [t, \cosh \phi].
\end{array}                    \label{geners}
\end{equation}
We normalize the Minkowski coordinates so that the commutator is
given by $[t^\prime, x^\prime] = i\kbar$. One finds that the
corresponding Kasner commutator is
\begin{equation}
[t,x] = i\kbar (1 + o((i\kbar)^2)).
\end{equation}
We shall use rather the form
\begin{equation}
[t,\phi] = i\kbar t^{-1}(1 + o((i\kbar)^2)).
\end{equation}

The frame is given by
\begin{equation}
\theta^0 = dt^\prime, \qquad \theta^1 = dx^\prime.
\end{equation}
We can rewrite it  in terms of $t$ and $x$ using the
change of variables.  It is of interest however to express also the
calculus  in terms of the Kasner moving frame; as we have
already noticed in Equation~(\ref{tlo}) it is not a
noncommutative frame since the frame rotation is local. We
designate it therefore $\eta^a$ and define as
\begin{equation}
\begin{array}{l}
\eta^{0} = \cosh \phi\, \theta^{0} -
\sinh \phi\, \theta^{1},  \\[2pt]
\eta^{1} = - \sinh \phi \,\theta^{0} + \cosh \phi \,\theta^{1}.
\end{array}                                                   \label{773} 
\end{equation}
Since from $[\phi, \theta^a] = 0$ we obtain $[\phi, \eta^a] = 0$,
the equation  (\ref{773}) can be easily inverted. To complete the definition
of the differential calculus we need the commutation relations
$[t, \eta^a]$. These can only be calculated perturbatively. To
lowest order one finds
\begin{eqnarray}
&&[t, \eta^0] =  -i\kbar t^{-1} \eta^1 +
\tfrac 32 (i\kbar)^2 t^{-3} \eta^0 ,\\[4pt]
&&[t, \eta^1] =  -i\kbar t^{-1} \eta^0 +
\tfrac 32 (i\kbar)^2 t^{-3} \eta^1.
\end{eqnarray}
To the same order, the transformation~(\ref{geners}) of the
generators reads
\begin{equation}
\begin{array}{l}
x^\prime =  \sinh \phi\, t + 
\frac 12 i\kbar \cosh\phi\, t^{-1} + 
\frac 14 (i\kbar )^2\sinh\phi\,t^{-3}, \\[4pt]
t^\prime = \cosh \phi\, t +\frac 12 i\kbar \sinh\phi\, t^{-1}
+\frac 14 (i\kbar )^2\cosh\phi\,t^{-3}.
\end{array}
\end{equation}
These commutation relations determine a noncommutative geometry which
is a natural extension of the flat Kasner geometry.

\initiate
\section{Fuzzy donut: momentum relations}   \label{Donut}

Let us examine now further properties of the module structure
defined by the differential~(\ref{diff}). The exterior product is
a map from the tensor product of two copies of the module of
1-forms into the module of 2-forms. We shall identify the latter
as a subset of the former and write the product as
\begin{equation}
\theta^a\theta^b =
P^{ab}{}_{cd}\theta^c\otimes\theta^d.                   \label{exterior}
\end{equation}
The $ P^{ab}{}_{cd}$ are complex numbers which satisfy the projector
condition and a hermiticity
condition~\cite{CerFioMad00a}. The basis 1-forms anticommute for
$P^{ab}{}_{ cd} = \frac12(\delta^a_c\delta^b_d-\delta^b_c\delta^a_d)$.
The exterior derivative of $\theta^a$ is a 2-form, so it can be
written as
\begin{equation}
d\theta^a = -\frac 12 C^a{}_{\ bc}\, \theta^b\theta^c.         \label{s-e2}
\end{equation}
The $C^a{}_{ bc}$ are called the structure elements. They can be chosen
to satisfy  $C^a{}_{ bc} = C^a{}_{de} P^{de}{}_{bc}$.

Impose now the condition $d^2 = 0$. It gives
\begin{equation}
0 = d(df) = d(-\left[ \theta, f\right] ) =
-\left[ d\theta +\theta^2,f\right],                \label{d2}
\end{equation}
so it implies that $d\theta +\theta^2$ commutes with all elements of the
algebra. Since $d\theta +\theta^2$ is a 2-form, in the frame basis
it can be written as
\begin{equation}
d\theta +\theta^2 = -\frac 12 K_{ab}\theta^a\theta^b         \label{*}
\end{equation}
where the elements $K_{ab} $ are complex numbers. One can impose
$K_{ab} P^{ab}{}_{ cd} = K_{cd}$.  A straightforward calculation shows
that
\begin{eqnarray}
&& d\theta = -dp_a \theta^a -p_a d\theta^a =
\left[p_b,p_a\right]\theta^b\theta^a +
\frac 12 p_aC^a{}_{bc}\theta^b\theta^c ,\ncr
&& \theta ^2 = p_ap_b\theta^a\theta^b,
\end{eqnarray}
and hence (\ref{*}) reduces to
\begin{equation}
( p_c p_b + \frac 12  C^a{}_{ bc}p_a + \frac 12  K_{bc})
\theta^b\theta^c = 0.
\end{equation}
This can be written as
\begin{equation}
2p_b p_a P^{ab}{}_{ cd} +p_a C^a{}_{cd}+K_{cd} = 0.      \label{MA}
\end{equation}
The relation $d(f\theta^a -\theta^a f) = 0$ written in terms of the
momenta gives further restrictions.  It reads
\begin{equation}
[ p_b\delta^a_c +p_c\delta^a_b+\frac 12  C^a{}_{b c}, f]
\theta^b\theta^c = 0
\end{equation}
which means
\begin{equation}
\left( C^a{}_{bc}+2p_b\delta^a_c +
2p_c\delta ^a_b - F^a{}_{bc}\right) \theta^b\theta^c = 0,       \label{CF}
\end{equation}
where $F^a{}_{bc}$ are complex numbers. Thus the structure
elements, defined in Equation~(\ref{s-e2}) are linear in the momenta.
It follows immediately that
\begin{equation}
e_a C^a{}_{bc} = 0. \label{con}
\end{equation}
This relation must be also satisfied in the commutative limit and
constitutes a constraint on the frame. The example of
Section~\ref{Rind} shows that this condition is not necessarily
sufficient.  A frame has four degrees of freedom in two dimensions.
The constraint subtracts one therefrom.  On the other hand having
chosen a calculus, the choice of frame is equivalent to a gauge
condition. This can be made more transparent if the momenta exist in
which case the gauge condition can be expressed as the
condition~(\ref{con}). The commutation relation~(\ref{cond}) can be
thought of also as a gauge condition since it is necessary for the
existence of the momenta; there remain hence $4-1-1 = 2$ degrees of
freedom. Combining (\ref{MA}) and (\ref{CF}) we obtain the relation
\begin{equation}
2 p_c p_d P^{cd}{}_{ab} -p_c  F^c{}_{ab} - K_{ab} = 0.   \label{quadratic}
\end{equation}
The coefficients in (\ref{quadratic}) are complex numbers. We see 
that the momentum generators $p_a$ satisfy a quadratic relation.

One can readily find the conjugate momenta for a family of 2-dim metrics
with one Killing vector.  We shall exhibit all possible choices 
which yield differential calculi based on inner derivations. As frame
we choose
\begin{equation}
\theta^0 = f(x) dt, \quad f > 0, \qquad \theta^1 = dx.      \label{frame}
\end{equation}
and we suppose that $J^{01} = J^{01}(x)$.  The frame relations can be  written as
\begin{equation}
\begin{array}{ll}
dx\, x = x\, dx, & \quad dx\, t = t\, dx, \\[4pt]
dt\, x =  x\, dt, &\quad dt\, t = \big(t + i\kbar F)dt.        \label{33}
\end{array}
\end{equation}
We have set, for convenience
\begin{equation}
F = J^{01}\frac{d}{dx} \log f.
\end{equation}
The differential structure of the algebra can be written as
\begin{equation}
(dx)^2 = 0,\quad dx\, dt = - dt\, dx,\quad
( dt)^2 = -\tfrac 12 i\kbar F^\prime dx\, dt
\end{equation}
or as the relations
\begin{eqnarray}
&&(\theta^1)^2 = 0,\qquad \theta^0 \theta^1 = -\theta^1 \theta^0, \\[4pt]
&&(\theta^0)^2 = \tfrac 12 i\kbar  f F^\prime \theta^0 \theta^1
= 2\ep  \theta^0 \theta^1
\end{eqnarray}
with
\begin{equation}
\epsilon = 4 \kbar f F^\prime .
\end{equation}
It follows from the frame properties that $\epsilon$ is a constant.

Suppose now that the dual momenta exist. The duality relations are
\begin{equation}
\begin{array}{ll}
[p_{0}, t] = f^{-1}, & [p_{0}, x] = 0, \\[2pt]
[p_{1}, t] = 0, & [p_{1}, x] = 1.
\end{array}
\end{equation}
These relations allow us to identify $p_1$ with the partial derivative
with respect to $x$. If $\phi= \phi(x)$ then
\begin{equation}
[p_1, \phi] =  [p_1,x] \partial_x \phi = \partial_x \phi.
\end{equation}
On the other hand, for $\phi =\phi(t,x)$ we can write to first order
\begin{equation}
[p_0, \phi] = [p_0,t] \partial_t \phi = f^{-1}\partial_t \phi.
\end{equation}
If we denote as before $[p_{0}, p_{1}] = L_{01}$, the Jacobi
identities imply the relations
\begin{equation}
\begin{array}{ll}
[p_0,J^{01}] = 0,  &\quad [p_1,J^{01} ] =0,   \\[4pt]
[t,L_{01}] = -  f^\prime  f^{-2}, &\quad [x,L_{01}] =0.
\end{array}                                              \label{RF1}
\end{equation}
One can conclude again that $J^{01}$ is constant
and also that $L_{01}$ is a function of $x$ alone. We set
$J^{01}=1$. It follows that, neglecting the integration constants,
the `Fourier transformation' between the position and momentum
generators is given by
\begin{equation}
p_0 = - \dfrac{1}{i\kbar} \int f^{-1}, \qquad
p_1 = - \dfrac{1}{i\kbar}\, t.                         \label{tort}
\end{equation}
Each of the pairs $(t,x)$ and $(p_0, p_1)$ generates the algebra.

The array $P^{ab}{}_{cd}$ we write as
\begin{equation}
P^{ab}{}_{cd} =  \tfrac 12 \delta^{[a}_{c}\delta^{b]}_d +
\ep Q^{ab}{}_{cd}
\end{equation}
In dimension two, if we assume that metric depends on $x$ that is
on $p_0$ only, we find that
\begin{equation}
P^{ab}{}_{cd} p_a p_b = \frac 12 [p_c, p_d] + \ep Q^{00}{}_{cd} p^2_0
\end{equation}
and therefore $L_{01}$ is given by
\begin{equation}
L_{01} = K_{01} + p_0  F^0{}_{01} - 2 \ep p_0^2 Q^{00}{}_{01}.
\end{equation}
The structure elements are given by
\begin{equation}
C^0{}_{01} = F^0{}_{01} - 4 \ep p_0 Q^{00}{}_{01}.
\end{equation}
Symmetry and reality of the product imply that $Q^{ab}{}_{cd}$ has 
non-vanishing elements:
\begin{equation}
Q^{10}{}_{00} = - Q^{01}{}_{00} = 1,\qquad
Q^{00}{}_{01} = - Q^{00}{}_{10} = 1.                \label{Qq}
\end{equation}
We set also
\begin{equation}
K_{01} = \frac 1{i\kbar J^{01}} = \frac 1{i\kbar}, \qquad 
F^0{}_{01} =  - ib\mu,
\end{equation}
while $C^0{}_{10}$ is determined by the constraint
\begin{equation}
C^0{}_{ab} P^{ab}{}_{01} = C^0{}_{01}, \qquad
C^0{}_{01} + C^0{}_{10} = -2\ep C^0{}_{00}.
\end{equation}
We have then finally the expressions
\begin{eqnarray}
&&L_{01} = (i\kbar)^{-1}(1 - b \mu^{-1}(\ep p_0) -
2\mu^{-2} (\ep p_0)^2), \\[8pt]
&&C^0{}_{01} = - ib \mu - 4\ep p_0,
\end{eqnarray}
and a differential equation
\begin{equation}
-\ep \frac{dp_0}{dx} = \mu^2 - 
\ep b \mu p_0 - 2 (\ep p_0)^2                             \label{equa}
\end{equation}
for $p_0$. There are three cases to be considered.

The simplest is the case with $\mu^2 \to \infty$. The equation reduces to
\begin{equation}
-i\kbar\,\dfrac{dp_0}{dx} = 1.                        \label{pp3}
\end{equation}
One finds the relations
\begin{equation}
i\kbar p_0 = - x, \qquad f(x) = 1.                   \label{S13}
\end{equation}
This is  noncommutative Minkowski space.

An equally degenerate case is the case with $\mu^2 \to \infty$ and with
$\epsilon b=c\mu$. Equation~(\ref{equa}) can be written in the form
\begin{equation}
- i\kbar\, \dfrac{dp_0}{dx} = 1 - ic p_0.                \label{pp2}
\end{equation}
One finds the solution
\begin{equation}
i p_0 = c^{-1}(e^{-\kbar^{-1} cx} - 1), \qquad 
f(x) =  e^{\kbar^{-1} c x}.                            \label{S12}
\end{equation}
The change of variables
\begin{equation}
t^\prime = 2t, \qquad \mu x^\prime = 2c^{-1}(e^{-c x}-1),
\end{equation}
transforms the algebra into the algebra of de~Sitter space analyzed in
Section~\ref{Lob}.

The case which interests us the most here is that with $\mu$ finite.
With $b=0$ the equation becomes
\begin{equation}
-\ep\, \frac{dp_0}{dx} = \mu^2 -2 (\ep p_0)^2.
\end{equation}
If we introduce the notation
\begin{equation}
\beta^2 = 2 \mu^2 > 0
\end{equation}
the equation becomes
\begin{equation}
\frac{1}{\beta} \frac{d}{dx} \left(-2 \ep\beta^{-1} p_0 \right)
= 1 - \left(- 2\ep \beta^{-1} p_0 \right)^2.
\end{equation}
The solution is given by
\begin{equation}
i\kbar p_0 = - \beta^{-1} \tanh (\beta x), \qquad
f(x) = \cosh^2 (\beta x).                     \label{S1}
\end{equation}
The function $F$ is
\begin{equation}
F = - 2i\beta^2 \kbar p_0 = 2\beta \tanh(\beta x).
\end{equation}
We find therefore the identity
\begin{equation}
F^\prime + F^2 = f^{-1} f^{\pprime} =
2\beta^2 (1 + \tanh^2 (\beta x)).                    \label{gauss}
\end{equation}
The frame~(\ref{frame}) is given by
\begin{equation}
\theta^0 = \cosh^2(\beta x)dt =
\tfrac 12 (1 + \cosh (2\beta x)) dt, \qquad
\theta^1 = dx .
\end{equation}
Frames of similar type have
appeared~\cite{LemSa94,GegKun97,GruKumVas02} in 2-dimensional dilaton
gravity theories.  In the commutative limit the connection and the
curvature which correspond to this frame are
\begin{equation}
\t{\omega}^0{}_1 = \t{\omega}^1{}_0 = F \t{\theta}^0, \qquad
\t{\Omega}^0{}_1 = \t{\Omega}^1{}_0 =
-(F^\prime +F^2)\theta^0\theta^1 = - f^{-1} f^\pprime
\t{\theta}^0\t{\theta}^1.                                \label{lcm}
\end{equation}
The solution is a completely regular manifold of Minkowski signature
which has the Rindler metric as singular limit.
In the limit $\beta\to 0$
\begin{equation}
i\kbar p_0 = - x, \qquad f = 1,
\end{equation}
and one finds Minkowski space.  In `tortoise' coordinate $x^*$,
\begin{equation}
x^* = \int \dfrac{dx}{f(x)} 
\end{equation}
the frame is given by
\begin{equation}
\theta^0 = \dfrac {1}{1-x^{*2}}\,  dt, \qquad
\theta^1 = \dfrac {1}{1-x^{*2}}\,  dx^*.
\end{equation}
From~(\ref{tort}) we see that $x^* = -i\kbar p_0$.

Under a Wick rotation
\begin{equation}
u=2i\beta x, \qquad v= t
\end{equation}
the frame~(\ref{frame}) becomes
\begin{equation}
\theta^0 = \tfrac 12 (1 + \cos u) dv, \qquad
\theta^1 = \dfrac{1}{2i\beta}\,du
\end{equation}
and the line element in the commutative limit has the form
\begin{equation}
ds^2 = \tfrac 14 (1 +\cos\t{ u})^2  d\t{v}^2 + 
\tfrac {1}{4}\beta^{-2} d\t{u}^2.
\end{equation}
This is the surface of the torus embedded in $\b{R}^3$:
\begin{equation}
\t{ x}=\tfrac 12 (1+\cos \t{u})\cos\t{ v},\quad
\t{y}= \tfrac 12 (1+\cos \t{u})\sin\t{ v},\quad
\t{ z} =\tfrac 1{2} \beta^{-1}\sin \t{u},
\end{equation}
and for this reason we call this metric the `fuzzy donut'.
It is a singular axially-symmetric surface of Gaussian curvature
\begin{equation}
\t{K} = 2\beta^2 (1 - \tan^2 \tfrac 12 \t{u}).
\end{equation}
The donut is defined by the coordinate range $0 \leq u \leq 2\pi$,
$0 \leq v \leq 2\pi$, with a singularity at the point $u=\pi$.
In spite of the singularity, the Euler characteristic is given by
\begin{equation}
e[\c{A}] = \frac{1}{4\pi} \epsilon_{ab} \int \t{\Omega}^{ab} =
 - \frac{1}{2\pi} \int \t{\Omega}^0{}_1 =
 - \frac{1}{2\pi} \int d\t{\omega}^0{}_1 = 0
\end{equation}
as it should be.  If we suppose the same domain in the Wick rotated
real-$t$ region, then
\begin{equation}
0 \leq x \leq \beta^{-1}\pi, \qquad
0 \leq t \leq 2\pi .
\end{equation}
As $\beta\to \infty$ the donut becomes more and more squashed,
and this domain becomes an elementary domain in the limiting Minkowski
space.

\initiate
\section{Noncommutative differential geometry}             \label{dg}

We have presented several noncommutative `blurings' of classical
geometries, all of which are of dimension two. We have concentrated
our attention on the new aspects of the noncommutative theory,
especially the plethora of differential calculi and the relation of
the geometry to the symplectic structures.  We have not, in fact,
introduced the metric, the connection or the curvature on the
noncommutative space.  This can be done by taking the commutative
limit and using the definition of a metric in terms of the frame. It
can also be done~\cite{Mad00c} before the limit is taken.  To complete
the analysis of the family of examples discussed in
Section~\ref{Donut}, we mention the linear connections, the metric  and
the curvature without defining them in the full rigor; for details we
refer to~\cite{BurMacMad04}. Note that when the momenta exist the
metric is given; otherwise there is a certain ambiguity which must be
determined by field equations.

To define a linear connection one needs a
`flip'~\cite{Mou95,DubMadMasMou96b}, 
\begin{equation}
\sigma(\theta^a \otimes \theta^b) = S^{ab}{}_{cd} \theta^c \otimes \theta^d,
\end{equation} 
which in the present notation is
equivalent to a 4-index set of complex numbers $S^{ab}{}_{cd}$ which
we can write as
\begin{equation}
S^{ab}{}_{cd} = \delta_c^b \delta_d^a + \ep T^{ab}{}_{cd}.
\end{equation}
 The covariant derivative is given by
\begin{equation}
D \xi = \sigma(\xi \otimes \theta) - \theta \otimes \xi .
\end{equation}
In particular
\begin{equation}
D \theta^a = -\omega^a{}_c\otimes\theta^c = - (S^{ab}{}_{cd} - 
\delta^b_c \delta^a_d) p_b \theta^c \otimes \theta^d = 
-\ep T^{ab}{}_{cd}  p_b \theta^c \otimes \theta^d ,
\end{equation}
so the connection-form coefficients are linear in the momenta
\begin{equation}
\omega^a{}_c = \omega^a{}_{bc}\theta^b = 
\ep p_d T^{ad}{}_{bc}\theta^b .                         \label{LeviC}
\end{equation}
On the left-hand side of the last equation is a quantity
$\omega^a{}_c$ which measures the variation of the metric; on the
right-hand side is the array $T^{ad}{}_{bc}$ which is directly related
to the anti-commutation rules for the 1-forms, and more important the
momenta $p_d$ which define the frame. As $\kbar \to 0$ the right-hand
side remains finite and
\begin{equation}
\omega^a{}_{c} \to \t{\omega}^a{}_{c}.
\end{equation}
The identification is only valid in the weak-field approximation. 
The connection is torsion-free if the components satisfy the constraint
\begin{equation}
\omega^a{}_{ef}P^{ef}{}_{bc} = \tfrac 12 C^a{}_{bc}.          \label{t-free}
\end{equation}

The metric is a map 
\begin{equation}
g:\; \Omega^1(\c{A}) \otimes \Omega^1(\c{A}) \rightarrow \c{A}.
 \end{equation}
Using the frame it is defined by
\begin{equation}
g(\theta^a\otimes\theta^b) = g^{ab},
\end{equation} 
and bilinearity of the metric implies that $g^{ab}$ are complex
numbers.  In the present formalism~\cite{Mad00c} the metric is `real'
if it satisfies the condition
\begin{equation}
\bar g^{ba} = S^{ab}{}_{cd}g^{cd}.               \label{metreal}
\end{equation} 
`Symmetry' of the metric can be defined either using the projection
\begin{equation}
P^{ab}{}_{cd}g^{cd} = 0,                     \label{metsymP}
\end{equation} 
or the flip
\begin{equation}
S^{ab}{}_{cd}g^{cd} = cg^{ab}.             \label{metsymS}
\end{equation} 
We usually take the frame to be orthonormal in the commutative limit,
therefore one can write the metric as
\begin{equation}
g^{ab} = \eta^{ab} +\ep h^{ab}.
\end{equation} 
In the linear approximation, the condition of the reality of the
metric becomes
\begin{equation}
h^{ab} + \bar{h}^{ab} = -
T^{ba}{}_{cd}\eta^{cd} .
\end{equation}
The connection is metric if 
\begin{equation}
\omega^{a}{}_{bc} g^{cd} +
\omega^{d}{}_{ce} S^{ac}{}_{bf}
 g^{fe} = 0,                                      
\end{equation}
or linearized,
\begin{equation}
T^{(ac}{}_d{}^{b)} = 0.                      
\end{equation}

In our 2-dim model the frame is of the form
\begin{equation}
\theta^0 = f(x) dt, \qquad
\theta^1 = dx.
\end{equation}
The torsion-free metric-compatible connection and the curvature are
classically given by the expressions~(\ref{lcm}). From these
expressions we see that the geometry is flat only if $f(x)$ is linear
in $x$. We recall that $\epsilon = \kbar\mu ^2$. To first order the
fuzzy calculus differs from the commutative limit in the two relations
\begin{eqnarray}
&&\theta^0 \theta^1 = P^{01}{}_{ab} \theta^a \theta^b = 
\tfrac 12 \theta^{[0} \theta^{1]} + \ep Q^{01}{}_{00} (\theta^0)^2
\nonumber\\[2pt] 
&&\phantom{\theta^0 \theta^1} =
\tfrac 12 \theta^{[0} \theta^{1]}  - \ep q  (\theta^0)^2   \label{17}  \\[4pt]
&&(\theta^0)^2  =  P^{00}{}_{ab} \theta^a \theta^b = \ep
Q^{00}{}_{01} \theta^{[0} \theta^{1]} = \ep q  \theta^{[0} \theta^{1]}.
\end{eqnarray}
These can be better written as
\begin{equation}
\theta^{(0} \theta^{1)} = -2 \ep q  (\theta^0)^2, \quad
(\theta^0)^2  = \ep q  \theta^{[0} \theta^{1]},                               
\end{equation}
and to first order reduce to 
\begin{equation}
\theta^{(0} \theta^{1)} = 0, \quad
(\theta^0)^2  = 2 \ep q  \theta^{0} \theta^{1}.                  \label{20} 
\end{equation}
The quantity $q$ which we have introduced in (\ref{17}-\ref{20}) is a
constant, $q =0$ in the cases of flat and de~Sitter noncommutative
spaces and $q=1$ in the fuzzy donut case. We will restrict our
considerations to the latter.

The differentials of the frame are given by
\begin{equation}
d\theta^0 = - C^0{}_{01} \theta^0 \theta^1, \qquad d\theta^1 = 0,
\end{equation}
with
\begin{equation}
C^0{}_{01} = - 4 \ep p_0 Q^{00}{}_{01} = - 4\ep p_0.
\end{equation}
The only non-vanishing components of the connection are
\begin{equation}
\omega^0{}_1 = \omega^1{}_0 = -4\ep p_0 \theta^0 = F \theta^0,
\end{equation}
and from (\ref{LeviC}) we  find 
\begin{equation}
T^{00}{}_{01} = T^{10}{}_{00} = -4.                     \label{t-us}
\end{equation}
To first order the condition that the torsion vanish is
the equation~(\ref{t-free}); it is satisfied by the values we obtain.
The curvature 2-form has components
\begin{eqnarray}
&&\Omega^0{}_1 = - (F^\prime + F^2) \,\theta^0 \theta^1\\[4pt]
&&\Omega^0{}_0 = \Omega^0{}_0 = 2\ep F^2 \theta^0 \theta^1.
\end{eqnarray}
Therefore to lowest order from (\ref{gauss}) we find the Gaussian curvature
\begin{equation}
\Omega^0{}_1 = - 2\beta^2 (1 + \tanh^2 (\beta x))\theta^0 \theta^1.
\end{equation}

We must define a `real', `symmetric' metric.
There are in principle four possible ways to define it
 depending on which of two possible ways one chooses to define
symmetry, and whether or not one includes a twist in the extension
of the metric to the tensor product.
In all cases the torsion-free condition yields the relation
\begin{equation}
T^{abcd} = 2(Q^{bcda}_- +Q^{bdca}_- +Q^{abdc}_-) ,   \label{TQ}
\end{equation}
and the reality of the metric 
\begin{equation}
h^{ab} + \bar h^{ab} = -
T^{ba}{}_{cd}\eta^{cd},
\end{equation}
both in the linear approximation. 
Here we denote $Q_-^{ab}{}_{cd} = \frac 12 Q^{ab}{}_{[cd]}$,  
$Q_+^{ab}{}_{cd} = \frac 12 Q^{ab}{}_{(cd)}$.
The projector $P^{ab}{}_{cd}$ is hermitean
if
\begin{equation}
Q_+^{abcd} = \pm Q_-^{cdab},
\end{equation}
with plus in the case of no twist and minus with twist. If one use the flip
to define symmetry, then  for some $\gamma$
the linearized perturbation must satisfy
\begin{equation}
h^{[ab]} = T^{ab}{}_{cd} \eta^{cd} - \gamma \eta^{ab}
\end{equation}
if the metric is to be symmetric.
If one use the product to define symmetry then
\begin{equation}
h^{[ab]} = - 2 Q_+^{abcd} \eta_{cd} .
\end{equation}
In the present example the only consistent choice is the 
following 
\begin{equation}
h^{[ab]} = - 2 Q_+^{abcd} \eta_{cd} = - 2 Q_-^{cdab} \eta_{cd}.
\end{equation}
Thus for the symmetric and real  metric we obtain
\begin{equation}
g^{ab} = \eta^{ab} + \ep h^{ab}, \qquad h^{ab} = 2
\left(\begin{array}{cc} 0 &1\\[2pt] 0 &0 \end{array}\right).
\end{equation}
The $\eta^{ab}$ here is the matrix of components of the canonical
Minkowski metric; to it can be added an antisymmetric real matrix which is not
fixed:
\begin{equation}
\eta^{ab} \mapsto \eta^{ab} +
\epsilon \left(\begin{array}{cc} 0 & a \\[2pt] -a &0 \end{array}\right).
\end{equation}
This ambiguity exists already at the classical level.

\initiate
\section{Higher-order effects}     \label{ho} 

To find the second order corrections to our system, we
 write
the 4-index tensors as matrices ordering the indices $(01,10,11,00)$.
Let $P_0$ and $S_0$ be respectively the canonical
projector and the flip
\begin{equation}
P_0=\left(\begin{array}{cccc}
1/2 & -1/2 & 0 & 0\\[2pt]
-1/2 & 1/2 & 0 & 0\\[2pt]
0 & 0 & 0 & 0\\[2pt]
0 & 0 & 0 & 0
\end{array}\right) ,\qquad
S_0=\left(\begin{array}{cccc}
0 & 1 & 0 & 0\\[2pt]
1 & 0 & 0 & 0\\[2pt]
0 & 0 & 1 & 0\\[2pt]
0 & 0 & 0 & 1
\end{array}\right) .
\end{equation} 
The projector constraints are, in matrix notation,
\begin{equation}
P^2 = P,\qquad   \bar  P \hat P = \hat P                  \label{p2p}
\end{equation}
where $\hat A^{ab}{}_{cd} = A^{ba}{}_{cd} = (S_0 A)^{ab}{}_{cd}.$ 
To lowest order these conditions become
\begin{equation}
\bar Q = Q, \qquad \hat Q_- = Q_-   .                  
\end{equation}
The twist constraints are
\begin{eqnarray}
&&\hat {\bar S} \hat S = 1, \\[4pt]          \label{S}
&&\hat S P + \bar P \hat P = 0, \\[4pt]      \label{SbarP}
&&SP + P = 0 .                                      \label{SP}
\end{eqnarray}
The last two identities are equivalent if
\begin{equation}
\bar P \hat P = \hat P.
\end{equation}
This condition was already imposed in (\ref{p2p}).

One can easily check that the first order solution of the previous
section which we write as $P = P_0 + \ep Q$, $S = S_0 + \ep T$, is
given by
\begin{equation}
Q = Q_- + Q_+,\qquad 
Q_- = \left(\begin{array}{cc}
0  &0 \\[2pt]
\tau  &0
\end{array}\right)   ,      \qquad
Q_+ = \left(\begin{array}{cc}
0  &-\tau^* \\[2pt]
0  &0
\end{array}\right),
\end{equation} 
and 
\begin{equation}
T = - 2 \left(\begin{array}{cc}
0 & \tau\tau^*\\[2pt]
 \tau\tau^*\sigma_1 &0 \end{array}\right)      .                
\end{equation}
We introduced the matrix $\tau$ and its transpose $\tau^*$:
\begin{equation}
\tau = \left(\begin{array}{cc}
0 &0\\[2pt]
1 &-1\end{array}\right),\qquad 
\tau \tau^* = 1-\sigma_3,\quad
\tau^* \tau = 1-\sigma_1.
\end{equation}

The constraints (\ref{p2p}), (\ref{S}-\ref{SP}) can be solved to
second order using inner automorphisms of the matrix algebra. Denote
\begin{eqnarray}
&&P = P_0 + \ep Q + (\ep)^2 Q_2,  \\[4pt]
&&S = S_0 + \ep T + (\ep)^2 T_2
\end{eqnarray}
 and introduce the automorphism
$P = W^{-1} P_0 W  $,
where $W$ is an arbitrary nonsingular $4\times 4$ matrix with 
inverse $W^{-1}$. We see immediately that
$P^2 = P $.
To satisfy the second condition of~(\ref{p2p}) on $P$ it is
sufficient to require that
\begin{equation}
\bar W S_0 = S_0 W     ,                           \label{barQS}
\end{equation}
and to recall  $\hat S = S_0 S$. Let
$W = \exp(\ep B)$.
To second order 
\begin{equation}
P = P_0 + \ep [P_0, B] + \tfrac 12 (\ep)^2 [[P_0,B],B],
\end{equation}
and the two expansions coincide if
\begin{equation}
Q = [P_0, B],     \qquad  Q_2 = \tfrac 12 [[P_0,B],B] = \tfrac 12 [Q,B].                \label{QmQp}
\end{equation}
It is easy to see that an appropriate solution is
\begin{equation}
B  = \left(\begin{array}{cc}
0 &-\tau^* \\[2pt] 
-\tau  &0
\end{array}\right).
\end{equation}
One can also check
\begin{equation}
P_0 B = Q_+, \quad BP_0 = - Q_-,
\quad
S_0 B = -BS_0 = - Q.
\end{equation}
The (\ref{barQS}) becomes the condition 
$\bar B S_0 = -S_0 B$, which in turn, since $B$ is real, is the
condition that $B$ and $S_0$ anticommute.  

The solution for $T_2$ is
\begin{equation}
 T_2 = \tfrac 12 T S_0 T. 
\end{equation}
To check whether the twist constraints hold,
introduce
\begin{equation}
A  = \left(\begin{array}{cc}
0 &\tau\tau^* \\[2pt] 
\tau\tau^*  &0
\end{array}\right)
,\qquad
T = - 2AS_0, \quad (A - B) P_0 = 0 .
\end{equation}
At least to second order we have
\begin{equation}
S = S_0 X, \quad S = Y S_0, \qquad 
X = \exp (\ep S_0T), \quad Y = \exp (\ep T S_0).
\end{equation}
The twist constraint
\begin{equation}
\hat{\bar S} \hat S = S_0 (S_0 \bar X) S_0 (S_0 X)
= S_0^2 \bar X X = 1
\end{equation}
follows. Further, consider the identity
\begin{eqnarray}
&&(1 - WYW) P_0 = 
(1 - \exp(\ep B) \exp(\ep TS_0) \exp(\ep B)P_0\nonumber\\[4pt]
&&
= - \ep (TS_0 + 2B) P_0 + \cdots
= \ep (T - 2B)P_0 + \cdots\nonumber\\[4pt]
&&
= \ep (T + 2Q)P_0 + \cdots
= (\ep)^2 H + o((\ep)^3)
\end{eqnarray}
with
\begin{equation}
H =\left(\begin{array}{cccc}
1  &-1  &0 &0\\[2pt]
-5  &5 &0 &0\\[2pt]
 0  &0  &0 &0\\[2pt]
 0  &0  &0 &0\\ \end{array}\right).
\end{equation}
To lowest order, therefore,
\begin{eqnarray}
&&SP+P  = Y S_0 W^{-1} P_0 W + P = YW S_0 P_0 W + P\nonumber\\[4pt]
&&\phantom{SP+P}= W^{-1}(1 -WYW) P_0 W = (\ep)^2 [H,B] + \cdots 
\end{eqnarray}
so all constraints are satisfied at least to second order.

The second-order metric is 
\begin{equation}
g = \sqrt{\bar X} g_0.
\end{equation}
To analyse the metric we write it as a 4-vector.
We see then that if $\bar g_0 = g_0$,
\begin{equation}
\hat{\bar g} = S_0 \sqrt{X} g_0 = S_0 X \sqrt{\bar X} g_0 = S g.
\end{equation}
The metric is real. Since also
\begin{equation}
P g = W^{-1} P_0 W \sqrt{\bar X} g_0 = W^{-1} [P_0, W \sqrt{\bar X}] g_0
\end{equation}
the metric is symmetric to the extent that
\begin{equation}
[P_0, W \sqrt{\bar X}] g_0 = 0.
\end{equation}
To first order this condition becomes
\begin{equation}
g = g_0 - \tfrac 12  \ep S_0 T g_0 =  = \left(\begin{array}{cc}
-1 &0 \\[2pt] 
0  &1
\end{array}\right) + 2\ep \left(\begin{array}{cc}
0 &1 \\[2pt] 
0  &0
\end{array}\right).
\end{equation}
We saw in Section~\ref{dg} that this metric is compatible with the
connection 
\begin{equation}
\omega^a{}_b = S^{ac}{}_{db}\theta^d p_c + \theta \delta^a_b, \qquad 
\theta = - p_a \theta^a
\end{equation}
to first order in the expansion parameter.

In general a connection is metric-compatible if the condition 
\begin{equation}
\omega^i{}_{kl} g^{lj} +
\omega^j{}_{ln} S^{il}{}_{km} g^{mn} = 0                     \label{metcom-c}
\end{equation}
is satisfied. This can be written in a more familiar form if one
introduce the `covariant derivative'
\begin{equation}
D_i X^j = \omega^j{}_{ik} X^k
\end{equation}
which is twisted:
\begin{equation}
D_i (X^j Y^k) = D_i X^j Y^k +S^{jl}{}_{im} X^m D_l Y^k.
\end{equation}
Condition~(\ref{metcom-c}) becomes then
\begin{equation}
D_i g^{jk} = 0.
\end{equation}
If $F^i{}_{jk} = 0$ then one can also express the condition as
\begin{equation}
S^{im}{}_{ln} g^{np} S^{jk}{}_{mp} = g^{ij} \delta^k_l.        \label{3.6.40}
\end{equation}
We have not succeeded in finding a connection which is
metric-compatible and tor\-sion-free to second order; there are, however,
solutions with torsion which are metric-compatible.

\initiate
\section{Conclusions}

Several models have been found which illustrate a close relation
between noncommutative geometry in its `frame-formalism' version and
classical gravity. Heuristically, but incorrectly, one can formulate
the relation by stating that gravity is the field which appears when
one quantizes the coordinates much as the Schr\"odinger wave function
encodes the uncertainty resulting from the quantization of phase
space.

The first and simplest of these is the fuzzy-sphere which is a
noncommutative geometry which can be identified with the 2-dimensional
(euclidean) `gravity' of the 2-sphere. The algebra in this case is an
$n\times n$ matrix algebra; if the sphere has radius $r$ then the
parameter $r/n$ can be interpreted as a lattice length. With the
identification this model illustrates how gravity can act as an
ultraviolet cutoff, a regularization which is very similar to the
`point splitting' technique which has been used when quantizing a
field in classical curved backgrounds. It can also be compared with
the screening of electrons in plasma physics, which gives rise to a
Debye length proportional to the inverse of the electron-number
density $n$. The analogous `screening' of an electron by virtual
electron-positron pairs is responsible for the reduction of the
electron self-energy from a linear to logarithmic dependence on the
classical electron radius.  Other models have been found which
illustrate the identification including an infinite series in all
dimensions.

In the present paper yet another model is given, one which although
representing a classical manifold of dimension 2 is of interest
because the classical `gravity' which arises has a varying Gaussian
curvature. The authors will leave to a subsequent article the delicate
task of explaining exactly which property of the metric makes it
`quantizable'.  This geometry could furnish a convenient model to
study noncommutative effects, for example in the colliding-$D$-brane
description of the Big-Bang proposed by Turok \& Steinhardt
\cite{TurSte04}. The 2-space describing the time evolution of the
separation of the branes has been shown to be conveniently described
using Rindler coordinates. One can blur this geometry by using the
metric and connection described here.  The flat geometry would have to
be replaced by the one given in this section; in the limit $q\to 0$ it
would become flat.

The donut example is of importance in that is is the first
explicit construction of an algebra and differential calculus
which is singularity-free in the Minkowksi-signature domain and
which has a non-constant curvature. There are two aspects of this
problem. To construct a classical manifold from a differential
calculus is relatively simple once one has constructed the frame.
One takes formally the limit and uses the so constructed moving
frame to define the metric. This is contained in the upper right of
the following little diagram
\begin{equation}
\begin{array}{ccc}
\buildrel{\mbox{Fuzzy}}\over{\mbox{Frame}}&\longrightarrow
&\buildrel{\mbox{Classical}}\over{\mbox{Frame}}\\[6pt]
\vert &&\vert\\[-5pt]
\downarrow &&\downarrow\\[6pt]
\buildrel{\mbox{\it Fuzzy}}\over{\mbox{\it Geometry}}&\longrightarrow
&\buildrel{\mbox{Classical}}\over{\mbox{Geometry}}
\end{array}
\end{equation}
More difficult is the construction of a `fuzzy geometry' which would
fill in the lower left of the diagram and would be such that the classical
geometry is a limit thereof. But this step is very important since it
gives an extension of the right-hand side into what could eventually
be a domain of quantum geometry. It is the box in the
to-be-constructed lower left corner where possibly
one can find an interesting extension of the metric containing
correction terms which describe the noncommutative structure.

We have not succeeded however to completely extend this geometry to
all orders in the noncommutativity parameter $\ep$. This will be
considered in a subsequent article. There is evidence that the
extension will involve a non-vanishing value of the torsion 2-form.
The metric is extended into the noncommutative domain so as to
maintain such formal properties as reality and symmetry. The
interpretation however as a length requires more attention when the
`coordinates' do not commute.

\section*{Acknowledgment}

Part of this work was done while the authors were visiting ESI in
Vienna.  They would like to thank the director for his hospitality as
well as T.~Grammatikopoulos J.~Mourad, T.~Sch\"ucker and G.~Zoupanos
for enlightening conversations.

\providecommand{\href}[2]{#2}\begingroup\raggedright\endgroup


\end{document}